\documentclass[preprint]{vgtc}                          

\graphicspath{{figures/}{pictures/}{images/}{./}} 

\usepackage{times}                     

\usepackage{tabu}                      
\usepackage{booktabs}                  
\usepackage{lipsum}                    
\usepackage{mwe}                       
\usepackage{amsmath}
\usepackage{algorithm}
\usepackage{algpseudocode}
\usepackage{enumitem}

\usepackage{mathptmx}                  

\onlineid{0}

\vgtccategory{Research}

\vgtcinsertpkg

\preprinttext{\parbox{0.9\textwidth}{This document is licensed under \href{https://creativecommons.org/licenses/by/4.0/}{CC-BY (Creative Commons Attribution 4.0 International)}. This paper was accepted by the 2024 IEEE VIS Workshop on Progressive Data Analysis and Visualization (PDAV).}}

\title{Progressive Glimmer: \\ Expanding  Dimensionality in Multidimensional Scaling}

\author{Marina Evers\thanks{e-mail: marina.evers@visus.uni-stuttgart.de} \\ \scriptsize University of Stuttgart \and David Hägele\thanks{e-mail: david.haegele@visus.uni-stuttgart.de} \\ \scriptsize University of Stuttgart \and Sören Döring
\thanks{e-mail: st142532@stud.uni-stuttgart.de} 
\\ \scriptsize University of Stuttgart \and Daniel Weiskopf\thanks{e-mail: daniel.weiskopf@visus.uni-stuttgart.de} \\ \scriptsize University of Stuttgart}

\teaser{
  \centering
  \includegraphics[width=\linewidth]{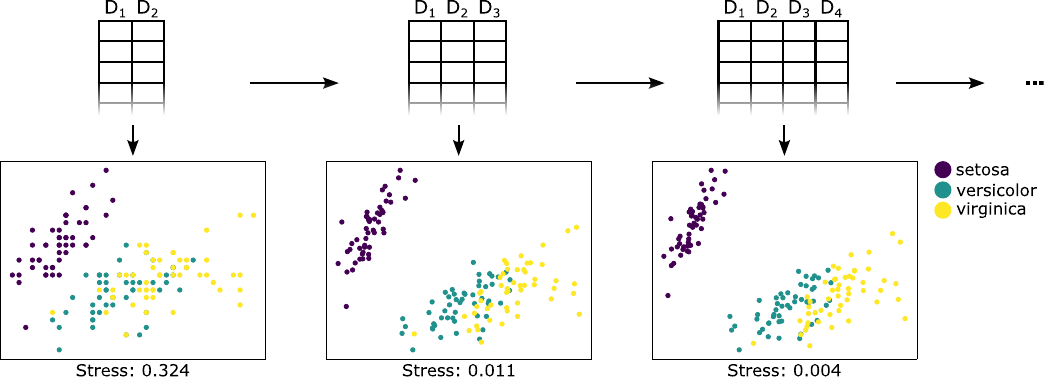}
  \caption{Progressive Glimmer supports updating the number of dimensions iteratively and performing consistent updates in the embedding. In this example, the algorithm is applied to the Iris dataset.}
  \label{fig:teaser}
}

\abstract{
Progressive dimensionality reduction algorithms allow for visually investigating intermediate results, especially for large data sets. While different algorithms exist that progressively increase the number of data points, we propose an algorithm that allows for increasing the number of dimensions. Especially in spatio-temporal data, where each spatial location can be seen as one data point and each time step as one dimension, the data is often stored in a format that supports quick access to the individual dimensions of all points. Therefore, we propose Progressive Glimmer, a progressive multidimensional scaling
(MDS) algorithm. We adapt the Glimmer algorithm to support progressive updates for changes in the data's dimensionality. We evaluate Progressive Glimmer's embedding quality and runtime. We observe that the algorithm provides more stable results, leading to visually consistent results for progressive rendering and making the approach applicable to streaming data. We show the applicability of our approach to spatio-temporal simulation ensemble data where we add the individual ensemble members progressively.
} 


\begin{document}


\firstsection{Introduction}

\maketitle

In this paper, we present a progressive version of the Glimmer algorithm~\cite{glimmer}, which is a dimensionality reduction (DR) method performing multidimensional scaling (MDS). 
Many DR methods, especially metric MDS, are prone to long runtimes.
Numerous approaches to improve efficiency have been proposed using hierarchies~\cite{vandermaaten2014barneshut}, interpolation~\cite{desilva2004landmark}, or multilevel processing as leveraged by Glimmer.
While these optimizations considerably reduce runtimes, the computations still take seconds to minutes, slowing down the analysis workflow as users need to wait for the computations to finish.
Progressive visual analytics~\cite{stolper2014progressiveVA} recommends using progressive methods that avoid disrupting the analysis process by providing intermediate results throughout the computation process.

Methods that iteratively refine their results, e.g., those leveraging gradient descent, naturally provide intermediate results after each iteration.
However, to perform those methods, all of the data has to be fully available.
In contrast, our progressive Glimmer allows for adding new data dimensions and updating the embedding efficiently without recomputation and, thus, supporting data chunking~\cite{ulmer2023survey}.
The algorithm can start on partially available data, e.g., when dimensions of the data set are located at different sources and introduce latency due to loading and preprocessing. It is also possible to limit the computation time for creating intermediate results.
Progressive Glimmer is also applicable in the related and often confused streaming setting, where new data continuously arrives.

Our contributions can be summarized as:
\begin{itemize}[noitemsep, topsep=1mm]
    \item An algorithm for progressively applying the Glimmer MDS algorithms on data with an increasing number of dimensions.
    \item An evaluation of the algorithm's performance and comparing the embedding quality to the non-progressive version.
    \item The application of Progressive Glimmer to a real-world dataset from the domain of climate science.
\end{itemize}

\section{Related Work}
Former works related to ours comprise special MDS techniques, DR methods for streaming data, and progressive DR methods.
For an introduction to MDS and historical overview, we point the reader to the survey of Saeed~et~al.~\cite{saeed2018survey-mds}. 
Brandes and Pich~\cite{Brandes2007progressiveMDS} proposed a progressive version of classical MDS inspired by landmark MDS~\cite{desilva2004landmark}, where a rough approximation of the final embedding is achieved with a small number of pivot points, which is then progressively increased.
Our algorithm, in contrast, is a metric MDS technique that supports progressive refinement through incremental extension of the set of data dimensions. 
This ability is different from out-of-sample extensions~\cite{bengio2003out-of-sample, taskin2019out-of-sample} that allow the insertion of new data points of the same dimensionality into an existing embedding.
Another progressive DR method by Pezotti~et~al.~\cite{pezzotti2017approximateTsne} enables t-SNE to be employed in progressive visual analytics and refinement is steerable by specifying areas of interest where approximations are swapped for exact computations.
Glimmer approximates close and distant relationships between data points throughout the minimization process.
While not being steerable, the approximations improve on each iteration.

DR methods for streaming applications provide embeddings for subsets of the data that are extended or updated upon newly arriving data.
\emph{STREAMIT}~\cite{alsakran2011streamit} is a visual analytics approach using metric MDS for text document visualization with a continuously running force simulation that allows inserting new documents.
It also allows interactively adapting the similarity measure between documents, resulting in a layout update.
Our approach uses a similar idea where introducing new dimensions implies a change in high-dimensional point distances, resulting in a progressive refinement of the inter-point similarity.
Apart from progressive visualization, our approach creates consistent visualization while progressing.
The mechanic proposed in \emph{temporal MDS}~\cite{jaeckle2016tmds} to select a subset of dimensions per time-step over a sliding window can also be applied in our case.
While \emph{temporal MDS} uses one-dimensional embeddings and a heuristic for flipping mirrored subsequent results, our method can provide temporally coherent n-dimensional embeddings by updating previous results.

\section{Algorithm}
We extend the Glimmer algorithm by Ingram~et~al.~\cite{glimmer} to enable updates to the data's dimensionality.
First, we recapitulate the original algorithm and then introduce our adaptation.

The original Glimmer algorithm uses a multilevel process. It starts by creating a layout with a small subset of points, then uses the layout as the basis for a larger subset on the next level until reaching the top level, which consists of all points.
Instead of using the full information of inter-point distances to determine the layout, a force-based approximation is used as proposed by Chalmers~\cite{chalmers1996layout}.
In each iteration of this algorithm, every point has a fixed-size set of random points to which the distances are considered. This set of points is used to approximate the stress and compute the gradients for optimization.
It is updated by swapping half of the points for different ones but retaining the closest ones.
By repeating this process in each iteration, the point sets converge to consist of the nearest neighbors (near sets) and constantly changing other points (far sets).
Due to the small number of inter-point distances that are computed in each iteration, a tremendous speedup is achieved compared to the exact MDS computation. 
However, testing for convergence is difficult since the stress is no longer steadily decreasing and is subject to noise.
Therefore, the stress values of the last $m$ iterations are used to smooth the stress using a windowed-sinc filter.

\begin{algorithm}[t]
\caption{Progressive Glimmer}\label{alg:pglimmer}
\begin{algorithmic}[1]
\State $X \gets$ data[$\text{dim}_1$~\dots~$\text{dim}_{\ell}$] 
\State $Y \gets$ data[$\text{dim}_1, \text{dim}_2$]
\State $\mathcal{N} \gets$ random set of k neighbors for each $x \in X$
\State \Call{Chalmers--MDS}{$X,Y,\mathcal{N}$}
\State extend $X$ with more dimensions, repeat previous step
\vspace{1mm}
\Procedure{Chalmers-MDS}{$X, Y, \mathcal{N}$}
    \State $\delta \gets$ initial forces set to 0
    \State $s$ = [\,]
    \While{has not converged}
        \State \Call{layout}{$X, Y, \mathcal{N}, \delta$}
        \State append current stress to $s$
        \State $\mathcal{N}' \gets$ random set of k/2 neighbors for each $x \in X$
        \State $\mathcal{N} \gets$ keep close neighbors, replace others by $\mathcal{N}'$ 
    \EndWhile
\EndProcedure
\vspace{1mm}
\Procedure{layout}{$X, Y, \mathcal{N}, \delta$}
    \State stress $\gets 0$
    \For{$i \in \{1~\dots~\text{len}(X)\}$}
        \State neighbors $\gets \mathcal{N}_i$
        \State $D_i \gets$ distances between $X$[i] and $X$[neighbors]
        \State $d_i \gets$ distances between $Y$[i] and $Y$[neighbors]
        \State stress $\gets$ stress $+ \lVert D_i - d_i\rVert$
        \State $\delta_i \gets$ update force with MDS gradient from $D_i$ and $d_i$
    \EndFor
    \State $Y \gets Y + \delta$
\EndProcedure
\end{algorithmic}
\end{algorithm}

For a progressive version of the Glimmer algorithm, we subsequently add individual dimensions as shown in \Cref{fig:teaser}. Progressing in the direction of dimensions instead of in the dimension of points is strongly motivated by investigating spatio-temporal data with dimensionality reductions, even though it is generally applicable to other data types. If dimensionality reduction should be applied to spatio-temporal data such as simulation data, the data is commonly stored as individual time steps. However, loading and processing entire datasets at once is often time-consuming. To avoid preprocessing the entire dataset, we aim for a progressive algorithm that directly makes use of the existing data structures and provides intermediate results based on a subset of time steps. 
\Cref{alg:pglimmer} shows the pseudo-code of our algorithm.  

Progressive Glimmer strongly builds on the results created with fewer dimensions. 
Most importantly, it uses the configuration of the previous step as an initial condition.
By assuming relatively small changes when adding additional dimensions, using the hierarchical multi-level approach for finding a good initial embedding is unnecessary.
Therefore, the \emph{Chalmers-MDS} procedure of \Cref{alg:pglimmer} is executed on all data points once instead of repeatedly on several hierarchy levels. 
Additionally, we assume that the nearest neighbors are reasonably well approximated which is why we use the neighbors ($\mathcal{N}$ in \Cref{alg:pglimmer})  of the previous computation as a starting point for the new computation. 
Of course, the nearest neighbors are updated during the iterative optimization. 
For the first progression step, an initial condition is required. 
Depending on the use case, we propose to either use two dimensions as a starting point or apply the original Glimmer algorithm to a selected subset of the data.

\begin{figure*}[t]
  \centering
  \includegraphics[width=\linewidth]{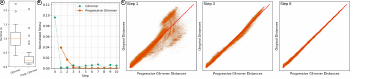}
  \caption{Runtime and stability of Glimmer and Progressive Glimmer. The computation time is substantially faster for progressive Glimmer (a), where the indicated times are for one step of progressive Glimmer. The stress over the number of included dimensions (b) reveals a sensitivity to the initial condition but also rapid improvements confirmed in the Shephard diagrams (c).}
  \label{fig:evaluation}
\end{figure*}

The original Glimmer algorithm uses a windowed-sinc filter to smooth the stress computation. The authors found that a filter length of $50$ yields good results. However, this requires at least $50$ iterations for early termination. Our algorithm might start with an initial configuration that is very close to the optimization target. Therefore, Progressive Glimmer should be able to terminate earlier. We achieve this by shortening the filter to only require $10$ iterations. 
If more iterations are required, the filter length is increased in steps of $10$ until the original filter is obtained again. To limit the latency between rendered results, the maximum number of iterations $z$ can be set to a user-defined value. For smaller progression steps, it is possible to visualize the output not only after processing a chunk of data but after a set of iterations.

\section{Evaluation}

\begin{figure}
  \centering
  \includegraphics[width=0.7\linewidth]{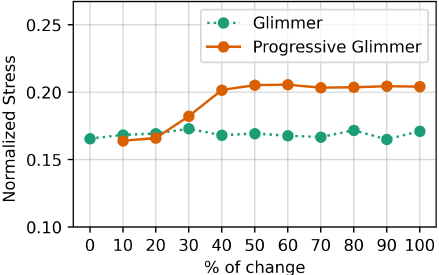}
  \caption{Comparing stress of Glimmer and Progressive Glimmer for different overlaps of a random data set. The data points of Glimmer are connected for better interpretability but are computed independently.}
  \label{fig:stepSize}
\end{figure}

In the following, we evaluate our approach with respect to stress and timing. As a reference, we use the original Glimmer algorithm applied to the subset of data. At first, we investigate the runtime shown in \Cref{fig:evaluation}a for the different runs of the algorithm on the spatio-temporal dataset discussed in \Cref{sec:usageScenarios}. To obtain the optimal result, the computation time is not limited by the number of iterations for this figure. While we observe several outliers, the mean runtime for Progressive Glimmer is substanitally smaller. The quality of Progressive Glimmer strongly depends on the embedding quality in the previous step. However, we observe an instability in this regard also in the original Glimmer algorithm. In \Cref{fig:evaluation}b, we show the change in stress over the different progressive steps. Here, the initial Glimmer application creates a relatively poor result as can also be seen in the Shepard diagram in \Cref{fig:evaluation}c. While the result of Progressive Glimmer in the first step could be substantially smaller, as shown by the stress value computed for the original Glimmer algorithm, the quality improves over the progressive steps. In this example, the stress values of Progressive Glimmer are already smaller than those of Glimmer after three steps. It is noteworthy that the stress decreases further and stays low consistently. The quality improvement can also be seen in the Shepard diagrams in \Cref{fig:evaluation}c. While these observations indicate a sensitivity to the initial conditions, they also show that a high-quality initial embedding is not required for good results after several progressive steps of the algorithm.

For each progressive step, we add a certain amount of new dimensions. To investigate the influence of the amount of newly added dimensions, we study the output quality based on the percentage of changed dimensions. Note that here we do not add new dimensions but move the data windows, which means that we remove the same number of dimensions that we add. While this is closer to a streaming paradigm than a progressive visualization, it provides clearly controlled results up to the point where the entire data is disjoint. The results are shown in \Cref{fig:stepSize}. For this evaluation, we used random data with 10,000 points because the complete lack of relation between the individual dimensions forms the worst case for our algorithm. Here, we identify that changing more than 20\% of the dimensions leads to a decrease in quality. However, note that especially spatio-temporal data, where time steps are used as dimension, contains smooth variations in time. This leads to smaller changes in the distances, and larger percentages of new dimensions can also lead to reasonably good results. Thus, the results of \Cref{fig:stepSize} present the worst case. While the worst-case is unlikely, it provides a lower bound for the amount of data that can be safely~added.

\section{Usage Scenarios}
\label{sec:usageScenarios}
In the following, we will show the applicability of our approach based on a real-world dataset. We use the Max Planck Institute for Meteorology Grand
Ensemble Simulations dataset (MPI-GE)~\cite{Maher2019} (available at https://esgfdata.dkrz.de/projects/mpi-ge/). In this paper, we use the temperature variations in the RCP8.5 scenario, which covers in total the time span from 2006 to 2100. With a spatial resolution of $96\times 192$, we obtain $18,432$ sample points that should be embedded. Applying dimensionality reduction to spatio-temporal data, where each point in the low-dimensional embedding corresponds to a spatial data point and the time steps are used as dimensions, allows for identifying regions with similar temporal behavior~\cite{evers2021uncertainty, evers2023envirvis}. First, we show the application to time-varying data, based on which we compare the effects of dealing with the dimensions in temporal order or using random access on the data. Second, we present an application to ensemble data. For this case, we use a similar concatenation of ensemble members as presented by Evers et al.~\cite{evers2021uncertainty} for the correlation computation. This usage scenario provides insights into adding larger chunks of data.

\textbf{Temporal Data.~~}
\begin{figure*}
  \centering
  \includegraphics[width=\linewidth]{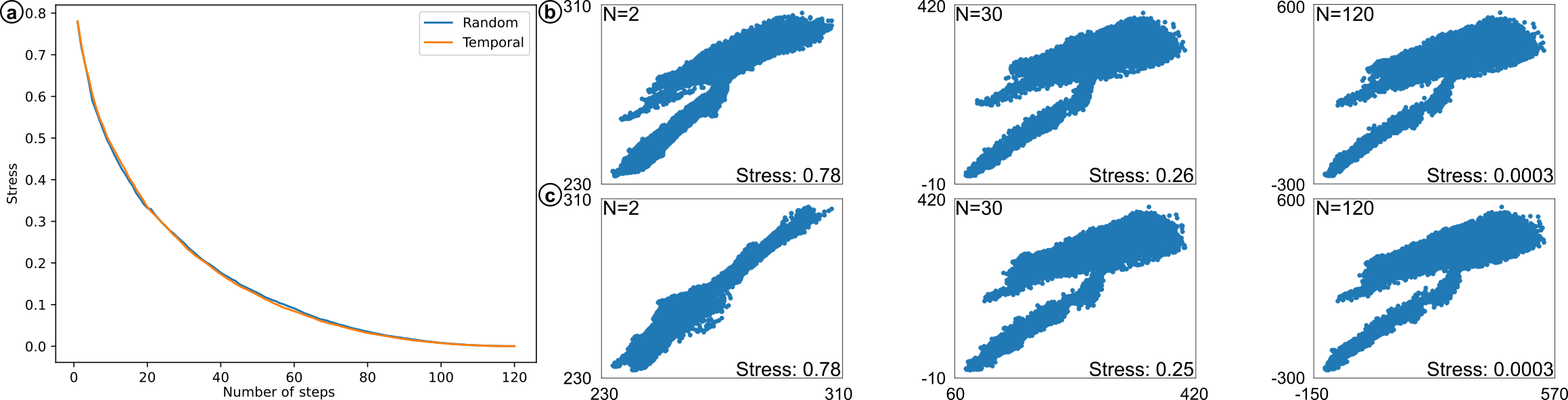}
  \caption{Progressive visualization for temporal data of one ensemble member of the MPI-GE dataset. The evolution of stress (a) does not show substantial differences between using temporal or random order when adding the time steps. The scatterplots for different numbers of time steps $N$ show that the variations for randomly adding dimensions (b) are small. When adding the steps in temporal order (c), the shape between the first two examples shown here varies substantially, while the following results are more similar.}
  \label{fig:temporal}
\end{figure*}
As a first usage scenario, we investigate a progressive MDS computation and visualization for temporal data. Here, we consider the first 10 years of the MPI-GE dataset, which corresponds to 120 timesteps, and use each time step as a dimension. We limit the computation time for each step by setting a threshold of $100$ iterations as a maximum. The results are shown in \Cref{fig:temporal}. Within this usage scenario, we also want to investigate the influence of the order in which the dimensions are added. Therefore, we compare adding individual dimensions in temporal order to adding the time steps in random order. 

The evolution of the stress when compared to the entire high-dimensional dataset is very similar for both cases (see \Cref{fig:temporal}a). In particular, in both cases, we observe a clear decrease indicating a convergence toward a result that would be obtained non-progressively. More interestingly, it is not clear which progression technique yields lower stress for intermediate results. Investigating the scatterplots reveals that the initial 2D embeddings, which are created by using the first two chosen dimensions as axes, show the largest differences between choosing a random order (see \Cref{fig:temporal}b) and the temporal order of the time steps (see \Cref{fig:temporal}c). When investigating the progression and the decrease in normalized stress, we see a constant decrease in stress that indicates that early stopping is not feasible without a loss in accuracy. However, the relative positions of the points remain stable, and mainly the scale of the point cloud changes for later progression steps. Thus, early termination can be feasible if primarily the shape of the embedding is of interest.

\begin{figure}
  \centering
  \includegraphics[width=\linewidth]{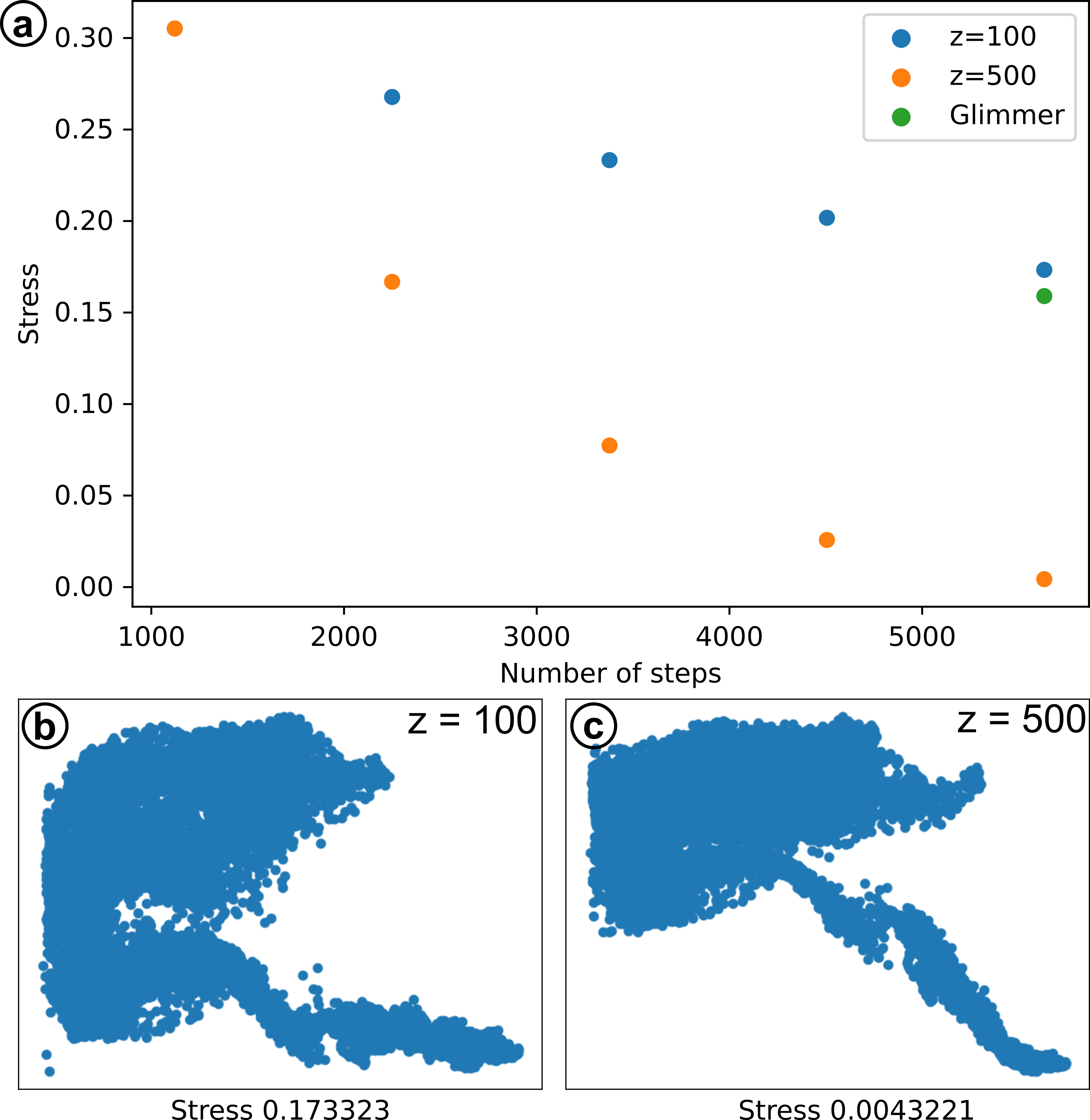}
  \caption{When adding the data in chunks of 1128 time steps (one ensemble member), more iterations are required to obtain smaller stress values (a). The result for five ensemble members for a maximum of 100 iterations (b) and a maximum of 500 iterations (c) show structural differences.}
  \label{fig:ensemble}
\end{figure}

\textbf{Ensemble Data.~~}
In the following, we investigate the application to ensemble data, where one data chunk corresponds to one ensemble member with $1128$ time steps and, thus, $1128$ dimensions. The results for five ensemble members are shown in \Cref{fig:ensemble}. For this case, we observe that limiting the maximum number of iterations to $z=100$ yields worse results than applying the algorithm directly. This observation can be explained by the relatively large changes in the distances and the low number of iterations to adapt. Relaxing the latency constraints and allowing for $z=500$ iterations for each data chunk substantially improves the result, as can also be seen in \Cref{fig:ensemble}b and c. For this case, even the computation using the original Glimmer algorithm is outperformed. Note that even for the larger number of iterations per data chunk, the latency can be limited further by rendering intermediate results after a user-defined number of iterations in the force-based optimization.

\section{Discussion and Conclusion}
In this paper, we presented Progressive Glimmer, which allows for computing MDS embeddings by progressively adding additional dimensions. While the algorithm can be applied to a wide variety of datasets, we see the main usage scenario in identifying features of spatio-temporal simulation data. We present evaluations regarding runtime and embedding quality measured by stress, as well as a usage scenario on a climate change dataset. In the taxonomy for progressive visualization~\cite{ulmer2023survey}, Progressive Glimmer applies data chunking and limits the latency by limiting the number of iterations per chunk of data. If even smaller update times are required, the intermediate results between the iterations of the force-based optimization algorithm can also be shown.

In this work, we used a CPU implementation. However, as Glimmer was designed to be executed on the GPU, we plan to incorporate the progressive computations in a GPU version, which can further improve runtime. In the future, we plan to evaluate the algorithm in more detail to optimize different hyper-parameters such as the number of neighbors and the convergence criterion for which we now mostly use the default values presented by Ingram et al.~\cite{glimmer}. One key point in progressive data visualization is the estimation of the quality. A common quality measure for MDS embeddings is the normalized stress. While the stress of the entire dataset provides very valuable quality insights, it is computationally expensive. Here, we plan to investigate the quality of different approximations that provide upper bounds also without the requirement to process the entire data. 
We plan to explore different applications such as the progressive computation of similarity images~\cite{evers2021uncertainty}. Based on the design of the algorithm, it can also be applied easily to streaming data 
and in the context of in-situ visualization. For example, the approach could be applied when calculating expensive numerical simulations to visualize intermediate results. Including steering options could lead to time savings in the data analysis process. In the future, we also plan to investigate the applicability in areas where consistent algorithms are required, such as for the computation of temporal MDS~\cite{jaeckle2016tmds}.

\newpage
\acknowledgments{Preliminary work for this paper is based on a master's thesis~\cite{thesisSoeren} by one of the co-authors, supervised by the other co-authors. Figures 2 and 3 have been adapted from the thesis. This work was funded by the Deutsche Forschungsgemeinschaft (DFG,
German Research Foundation)—Project ID 251654672—TRR 161
(Project A01).}

\bibliographystyle{abbrv-doi}

\bibliography{template}

\begin{thebibliography}{10}

\bibitem{alsakran2011streamit}
J.~Alsakran, Y.~Chen, Y.~Zhao, J.~Yang, and D.~Luo.
\newblock {STREAMIT}: Dynamic visualization and interactive exploration of text streams.
\newblock In {\em 2011 IEEE Pacific Visualization Symposium}, pp. 131--138, 2011. doi: {{%
10\hspace{.1pt}\discretionary{.}{%
}{.}\hspace{.4pt}1109\discretionary{/}{%
}{/}PACIFICVIS\hspace{.1pt}\discretionary{.}{%
}{.}\hspace{.4pt}2011\hspace{.1pt}\discretionary{.}{%
}{.}\hspace{.4pt}5742382}}


\bibitem{bengio2003out-of-sample}
Y.~Bengio, J.-F. Paiement, P.~Vincent, O.~Delalleau, N.~Roux, and M.~Ouimet.
\newblock Out-of-sample extensions for {LLE}, {Isomap}, {MDS}, {Eigenmaps}, and spectral clustering.
\newblock In S.~Thrun, L.~Saul, and B.~Sch\"{o}lkopf, eds., {\em Advances in Neural Information Processing Systems}, vol.~16, pp. 177--184. MIT Press, 2003.

\bibitem{Brandes2007progressiveMDS}
U.~Brandes and C.~Pich.
\newblock Eigensolver methods for progressive multidimensional scaling of large data.
\newblock In M.~Kaufmann and D.~Wagner, eds., {\em Graph Drawing}, Lecture Notes in Computer Science, pp. 42--53. Springer, Berlin, Heidelberg, 2007. doi: {{%
10\hspace{.1pt}\discretionary{.}{%
}{.}\hspace{.4pt}1007\discretionary{/}{%
}{/}978\discretionary{%
}{-}{-}3\discretionary{%
}{-}{-}540\discretionary{%
}{-}{-}70904\discretionary{%
}{-}{-}6\_6}}


\bibitem{chalmers1996layout}
M.~Chalmers.
\newblock A linear iteration time layout algorithm for visualising high-dimensional data.
\newblock In {\em Proceedings of Seventh Annual IEEE Visualization '96}, pp. 127--131, 1996. doi: {{%
10\hspace{.1pt}\discretionary{.}{%
}{.}\hspace{.4pt}1109\discretionary{/}{%
}{/}VISUAL\hspace{.1pt}\discretionary{.}{%
}{.}\hspace{.4pt}1996\hspace{.1pt}\discretionary{.}{%
}{.}\hspace{.4pt}567787}}


\bibitem{desilva2004landmark}
V.~De~Silva and J.~B. Tenenbaum.
\newblock Sparse multidimensional scaling using landmark points.
\newblock Technical report, Stanford University, 2004.

\bibitem{thesisSoeren}
S.~Döring.
\newblock Progressive multidimensional scaling for large data.
\newblock Master's thesis, University of Stuttgart, 2024.

\bibitem{evers2023envirvis}
M.~Evers, M.~Böttinger, and L.~Linsen.
\newblock Interactive visual analysis of regional time series correlation in multi-field climate ensembles.
\newblock In S.~Dutta, K.~Feige, K.~Rink, and D.~Zeckzer, eds., {\em Workshop on Visualisation in Environmental Sciences (EnvirVis)}. The Eurographics Association, 2023. doi: {{%
10\hspace{.1pt}\discretionary{.}{%
}{.}\hspace{.4pt}2312\discretionary{/}{%
}{/}envirvis\hspace{.1pt}\discretionary{.}{%
}{.}\hspace{.4pt}20231108}}


\bibitem{evers2021uncertainty}
M.~Evers, K.~Huesmann, and L.~Linsen.
\newblock Uncertainty-aware visualization of regional time series correlation in spatio-temporal ensembles.
\newblock {\em Computer Graphics Forum}, 40(3):519--530, 2021.

\bibitem{glimmer}
S.~Ingram, T.~Munzner, and M.~Olano.
\newblock Glimmer: {M}ultilevel {MDS} on the {GPU}.
\newblock {\em IEEE Transactions on Visualization and Computer Graphics}, 15(02):249--261, 2009. doi: {{%
10\hspace{.1pt}\discretionary{.}{%
}{.}\hspace{.4pt}1109\discretionary{/}{%
}{/}TVCG\hspace{.1pt}\discretionary{.}{%
}{.}\hspace{.4pt}2008\hspace{.1pt}\discretionary{.}{%
}{.}\hspace{.4pt}85}}


\bibitem{jaeckle2016tmds}
D.~Jäckle, F.~Fischer, T.~Schreck, and D.~A. Keim.
\newblock Temporal {MDS} plots for analysis of multivariate data.
\newblock {\em IEEE Transactions on Visualization and Computer Graphics}, 22(1):141--150, 2016. doi: {{%
10\hspace{.1pt}\discretionary{.}{%
}{.}\hspace{.4pt}1109\discretionary{/}{%
}{/}TVCG\hspace{.1pt}\discretionary{.}{%
}{.}\hspace{.4pt}2015\hspace{.1pt}\discretionary{.}{%
}{.}\hspace{.4pt}2467553}}


\bibitem{Maher2019}
N.~Maher, S.~Milinski, L.~Suarez-Gutierrez, M.~Botzet, M.~Dobrynin, L.~Kornblueh, J.~Kröger, Y.~Takano, R.~Ghosh, C.~Hedemann, C.~Li, H.~Li, E.~Manzini, D.~Notz, D.~Putrasahan, L.~Boysen, M.~Claussen, T.~Ilyina, D.~Olonscheck, T.~Raddatz, B.~Stevens, and J.~Marotzke.
\newblock The {M}ax {P}lanck {I}nstitute {G}rand {E}nsemble: {E}nabling the exploration of climate system variability.
\newblock {\em Journal of Advances in Modeling Earth Systems}, 11(7):2050--2069, 2019. doi: {{%
10\hspace{.1pt}\discretionary{.}{%
}{.}\hspace{.4pt}1029\discretionary{/}{%
}{/}2019MS001639}}


\bibitem{pezzotti2017approximateTsne}
N.~Pezzotti, B.~P.~F. Lelieveldt, L.~van~der Maaten, T.~Höllt, E.~Eisemann, and A.~Vilanova.
\newblock Approximated and user steerable {tSNE} for progressive visual analytics.
\newblock {\em IEEE Transactions on Visualization and Computer Graphics}, 23(7):1739--1752, 2017. doi: {{%
10\hspace{.1pt}\discretionary{.}{%
}{.}\hspace{.4pt}1109\discretionary{/}{%
}{/}TVCG\hspace{.1pt}\discretionary{.}{%
}{.}\hspace{.4pt}2016\hspace{.1pt}\discretionary{.}{%
}{.}\hspace{.4pt}2570755}}


\bibitem{saeed2018survey-mds}
N.~Saeed, H.~Nam, M.~I.~U. Haq, and D.~B. Muhammad~Saqib.
\newblock A survey on multidimensional scaling.
\newblock {\em ACM Computing Surveys}, 51(3), 2018. doi: {{%
10\hspace{.1pt}\discretionary{.}{%
}{.}\hspace{.4pt}1145\discretionary{/}{%
}{/}3178155}}


\bibitem{stolper2014progressiveVA}
C.~D. Stolper, A.~Perer, and D.~Gotz.
\newblock Progressive visual analytics: User-driven visual exploration of in-progress analytics.
\newblock {\em IEEE Transactions on Visualization and Computer Graphics}, 20(12):1653--1662, 2014. doi: {{%
10\hspace{.1pt}\discretionary{.}{%
}{.}\hspace{.4pt}1109\discretionary{/}{%
}{/}TVCG\hspace{.1pt}\discretionary{.}{%
}{.}\hspace{.4pt}2014\hspace{.1pt}\discretionary{.}{%
}{.}\hspace{.4pt}2346574}}


\bibitem{taskin2019out-of-sample}
G.~Taşkin and M.~M. Crawford.
\newblock An out-of-sample extension to manifold learning via meta-modeling.
\newblock {\em IEEE Transactions on Image Processing}, 28(10):5227--5237, 2019. doi: {{%
10\hspace{.1pt}\discretionary{.}{%
}{.}\hspace{.4pt}1109\discretionary{/}{%
}{/}TIP\hspace{.1pt}\discretionary{.}{%
}{.}\hspace{.4pt}2019\hspace{.1pt}\discretionary{.}{%
}{.}\hspace{.4pt}2915162}}


\bibitem{ulmer2023survey}
A.~Ulmer, M.~Angelini, J.-D. Fekete, J.~Kohlhammer, and T.~May.
\newblock A survey on progressive visualization.
\newblock {\em IEEE Transactions on Visualization and Computer Graphics}, 2023.

\bibitem{vandermaaten2014barneshut}
L.~van~der Maaten.
\newblock Accelerating t-{SNE} using tree-based algorithms.
\newblock {\em Journal of Machine Learning Research}, 15(93):3221--3245, 2014.

\end{thebibliography}
\end{document}